\documentclass[prl,twocolumn,amsmath,showpacs,
]{revtex4}
\usepackage{graphicx}

\newcommand{\beq}{\begin{equation}}
\newcommand{\eeq}{\end{equation}}
\newcommand{\bea}{\begin{eqnarray}}
\newcommand{\eea}{\end{eqnarray}}
\newcommand{\bml}{\begin{subequations}}
\newcommand{\eml}{\end{subequations}}
\newcommand{\ba}{\begin{array}}
\newcommand{\ea}{\end{array}}
\newcommand{\bc}{\begin{center}}
\newcommand{\ec}{\end{center}}
\newcommand{\commentout}[1]{{}}

\newcommand{\half}{\hbox{$\frac{1}{2}$}}

\newcommand{\eq}[1]{(\ref{#1})}
\newcommand{\etal} {{\it et al.\/}}

\newcommand{\vol}[1]{{\bf #1}}
\newcommand{\comment}[1]{{}}

\begin{document}

\title{Creating a Quantum Degenerate Gas of Stable Molecules via Weak Photoassociation}
\author{Matt Mackie and Pierre Phou}
\affiliation{Department of Physics, Temple University, Philadelphia, PA 19122}
\date{\today}

\pacs{03.75.Nt, 34.50.Rk, 42.50.Ct}

\begin{abstract}

Quantum degenerate molecules represent a new paradigm for fundamental studies and practical applications. Association of already quantum degenerate atoms into molecules provides a crucial shortcut around the difficulty of cooling molecules to ultracold temperatures. Whereas association can be induced with either laser or magnetic fields, photoassociation requires impractical laser intensity to overcome poor overlap between the atom pair and molecular wavefunctions, and experiments are currently restricted to magnetoassociation. Here we model realistic production of a quantum degenerate gas of stable molecules via two-photon photoassociation of Bose-condensed atoms. An adiabatic change of the laser frequency converts the initial atomic condensate almost entirely into stable molecular condensate, even for low-intensity lasers. Results for dipolar LiNa provide an upper bound  on the necessary photoassociation laser intensity for alkali-metal atoms $\sim30$~W/cm$^2$, indicating a feasible path to quantum degenerate molecules beyond magnetoassociation.

\end{abstract}

\maketitle

{\em Introduction.}--Quantum degenerate molecules offer a new level of control over experiments in ultracold chemistry~\cite{KRE10}, which could facilitate investigations in physical chemistry, e.g., the effect of long-range van der Waals forces on molecular dynamics~\cite{SKO99}, the role of fine and hyperfine interactions in chemical reactions~\cite{GAR08}, chemical reactions in reduced dimensions~\cite{ZLI09}, as well as ultraselective photodissociation and bimolecular reactions~\cite{MOO02}. Moreover, ultracold molecules could test the standard model and beyond, since quantum degeneracy allows for precise spectroscopic determination physical constants~\cite{HUD06}, and coupled atom-molecule systems are an analogue for quantum gravity~\cite{NAT10}.  Dipolar molecules in particular, with their controllable long range anisotropic interactions, are not only an excellent candidate for implementing quantum computation schemes~\cite{DEM02}, but allow for proxy investigations of exotic condensed-matter phases~\cite{BAR08}.

Since buffer gas cooling~\cite{WEI98} and electromagnetic deceleration~\cite{BET03} of molecules have yet to reach the ultracold regime, experiments have turned to association of already quantum degenerate atoms. Association of atoms into molecules occurs in the presence of a magnetic field due to the hyperfine interaction~\cite{TIM99}, or in the presence of a laser field due to the electric dipole interaction~\cite{WEI99}. For typical distances between quantum degenerate atoms, the magnetic hyperfine coupling between atoms and molecules can be much stronger than the electric dipole coupling at practical laser intensity ($\alt10$~W/cm$^2$). Associative production of a quantum degenerate gas of stable molecules via laser resonances, or photoassociation, is therefore complicated by the need for large intensity~\cite{WYN00,WIN05}, and experimental schemes have focused on magnetic resonances, or magnetoassociation~\cite{NI08,DAN09}. However, magnetic resonances are not always convenient or existent. Moreover, photoassociation would allow, for example, the ready possibility of phase imprinting the product molecules, or transferring angular momentum from the lasers to the molecules.

Photoassociation creates molecules in an electronically-excited state, so at least one additional laser is required to create stable molecules, in which case we have a two-photon transition. There are three distinct two-photon schemes for creating stable molecules: (i) Lasers with both intensity and frequency fixed, which leads to oscillations between atoms and stable molecules similar to Josephson oscillations in a superconductor. (ii) Lasers with pulsed intensity and fixed frequency, where atoms are converted into ground state molecules without ever creating electronically-excited molecules, i.e., the excited state is ``dark". (iii) Lasers with fixed intensity and slowly changing frequency, whereby the system adiabatically follows the ground state as it evolves from atoms into stable molecules. Josephson-like \cite{DRU99} and dark-state~\cite{DRU02} schemes have been reported previously, and require impractical laser intensity to overcome shifts of laser resonance due to elastic collisions between the particles. Adiabatic following is used extensively in magnetoassociation experiments~\cite{NI08,DAN09,KOH06}, but has somehow been overlooked as a means to create stable molecules with photoassociation. Here we report that, for practical laser intensities, a quantum degenerate gas of stable molecules can be created with two-photon photoassociation using adiabatic following.
\begin{figure}[t]
\centering
\includegraphics[width=7.75cm]{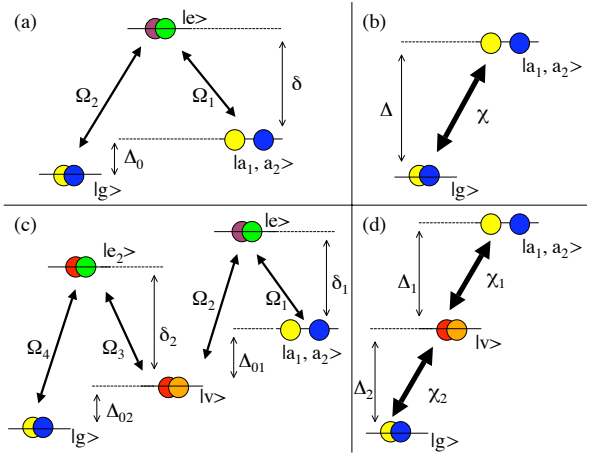}
\caption{(color online)~(a)~Two-photon photoassociation of atoms directly into stable molecules. (b)~If laser~1 is off resonance, a two-level system emerges. (c) Four-photon transitions through an intermediate vibrational state. (d) If lasers~1~and~3 are off-resonant, a three-level system emerges.}
\label{FEWL}
\end{figure}

{\em Basic Model.}--Without loss of generality, we focus on the creation of dipolar molecules via heteronuclear photoassociation~\cite{SHA99}. Consider a mixture of two atomic species that have Bose-condensed into the state $|a_1,a_2\rangle$. In a two-laser scheme [Fig.~\ref{FEWL}(a)], a photoassociation laser converts an atom from each species into an electronically-excited molecule in the state $|e\rangle$, where $\Omega_1$ is the atom-molecule coupling arising from the the electric dipole interaction, and $\delta$ is the one-photon detuning arising from the energy mismatch between the laser~1 photons and the energy of the molecular state $|e\rangle$ relative to the atomic state $|a_1,a_2\rangle$. The state $|e\rangle$ undergoes spontaneous radiative decay at the rate $\Gamma_0$, as well as stimulated photodissociative decay that will be introduced momentarily. A second laser converts excited molecules in $|e\rangle$ into stable molecules in the state $|g\rangle$, with $\Omega_2$ the molecule-molecule coupling due to the electric dipole interaction, and $\Delta_0$ the two-photon detuning that describes the energy mismatch between the total energy of two laser photons and the energy of the molecular state $|g\rangle$ relative to the atomic state $|a_1,a_2\rangle$. Elastic collisions between particles change the relative energy between the states, and therefore the detuning of the lasers, according to the coupling $\Lambda_{ij}\propto\rho a_{ij}/\mu_{ij}$, where $\rho$ is the total particle density, $a_{ij}$ is the $s$-wave scattering length, and $\mu_{ij}$ is the reduced mass for the $i$th and $j$th particles. Trapping of the particles and dipole-dipole interactions are negligible on the timescale for atom-molecule conversion. We return to the four laser scheme [Figs.~\ref{FEWL}(c,d)] in a moment.

When the photoassociation laser is far off resonance with the transition $|a_1,a_2\rangle\leftrightarrow|e\rangle$, i.e., for $|\delta|\gg\Gamma_0$, then one-photon transitions to the excited state $|e\rangle$ are negligible, and two-photon transitions directly to the stable molecular state $|g\rangle$ will dominate, as per Fig.~\ref{FEWL}(b). The two-photon coupling between the atoms and stable molecules is $\chi=\Omega_1\Omega_2/\delta$, the effective two-photon detuning $\Delta=\Delta_0-\Omega_2^2/\delta$, and the effective spontaneous decay rate $\Gamma=(\Omega/\delta)^2\Gamma_0$. The mean-field equations of motion for this system are
\bml
\bea
i\dot{a}_1 &=& \omega_1 a_1 -\half\chi a_2^* g,
\label{A1_EQM}
\\
i\dot{a}_2 &=& \omega_2 a_2 -\half\chi a_1^* g,
\label{A2_EQM}
\\
i\dot{g} &=&  (-\Delta+\omega_3-i\Gamma/2) g -\half\chi a_1a_2
\nonumber\\&&
  -\frac{\chi}{8\pi^2\omega_\rho^{3/2}}\!\int\!d\varepsilon\sqrt{\varepsilon}\,f(\varepsilon)A(\varepsilon),
\label{G_EQM}
\\
i\dot{A}(\varepsilon) &=& \varepsilon A(\varepsilon)
    -\chi\,f(\varepsilon)g.
\label{A_EQM}
\eea
\label{EQM}
\eml

Here  $a_{1(2)}$ is the probability amplitude for atomic species 1(2), $g$ is the amplitude for stable molecules, $A(\varepsilon)$ is the amplitude for atom pairs that are photodissociated out of the condensate, $\hbar\varepsilon=p^2/2\mu_{12}$ is the kinetic energy of a photodissociated pair, and $f(\hbar\varepsilon)$ is the energy spectrum of dissociated pairs. The role of photodissociation to noncondensate pairs is expected~\cite{JAV02} to be significant when $\chi\agt\omega_\rho$, where $\omega_\rho=\hbar\rho^{2/3}/2\mu_{12}$. Finally, the shifts in the energy levels due to elastic collisions between the particles are accounted for by $\omega_1=\Lambda_{11}|a_1|^2+\Lambda_{12}|a_2|^2+\Lambda_{13}|g|^2$, $\omega_2=\Lambda_{12}|a_1|^2+\Lambda_{22}|a_2|^2+\Lambda_{23}|g|^2$, and $\omega_3=\Lambda_{13}|a_1|^2+\Lambda_{23}|a_2|^2+\Lambda_{33}|g|^2$.

{\em Explicit Parameters.}--We focus on LiNa since it is the lightest (alkali-metal) heteronuclear molecule, and therefore provides an upper bound on the required photoassociation laser intensity. The atom-molecule coupling $\Omega_1$ for the $|a_1,a_2\rangle\leftrightarrow|e\rangle$ transition is obtained by scaling a typical $\Omega_1$ for homonuclear photoassociation of $^7$Li~\cite{MAC08}, so that $\Omega_1/2\pi=\sqrt{(\mu_0/\mu_{12})(\rho/\rho_0)(I/I_0)}\times290$~kHz, where $\mu_0=3.5$~a.u., $\rho_0=4\times10^{12}$cm$^{-3}$, and $I_0=28$~W/cm$^2$. For $\Gamma_0/2\pi=12$~MHz, the detuning of the photoassociation laser is set to $\delta=10^2\Gamma_0$, which is large enough suppress spontaneous decay with $\chi=100\Gamma$ for reasonable intensities, but small enough to avoid interference from neighboring states of the photoassociation target $|e\rangle$.  The molecule-molecule coupling is set to $\Omega_2\alt\Gamma_0$, which is generally possible for reasonable laser intensity.  For dense condensates, $\rho=10^{14}$~cm$^{-3}$, the largest coupling for elastic collisions is estimated~\cite{PEL08} as $\Lambda_{\rm max}\approx0.3\omega_\rho$, so that strong with respect to two-photon photodissociation is essentially the same as strong with respect to collisions. As per Table~\ref{PARAMS}, we consider strong, moderate and weak couplings $\chi$, although only weak coupling corresponds to photoassociation laser intensities that are easily achieved~\cite{MCK02}.

\begin{table}[b]
\caption{Estimated typical parameters for $^7$Li-Na.}
\begin{ruledtabular}
\begin{tabular}{cccc}
$\chi$ [$\omega_\rho$] & $\Omega_1$ [$\omega_\rho$] & $I_1$ [W/cm$^2$] 
  & $\Omega_2$ [$\Gamma_0$]\\
\hline \vspace{-0.125cm}\\
6.16 & 602 & 3.05$\times10^3$ &  1.02\\
0.617 &  191 & 307 & 0.324\\
0.065 & 62 & 32.3 & 0.105\\
\vspace{-0.25cm}
\label{PARAMS}
\end{tabular}
\end{ruledtabular}
\end{table}

{\em Adiabatic Following.}--Instead of the mismatch between the two-photon photon energy and the relative atom-molecule energy, think of the detuning $\Delta$ as simply the energy of the stable molecular state--relative to the atomic state--that is tunable according to the frequency of the lasers. For a laser frequency such that $\Delta>0$, as in Fig.~\ref{FEWL}(b), the ground state of the system is then stable molecules, whereas the ground state of the system is atoms for $\Delta<0$. According to the adiabatic theorem~\cite{TON10}, if the system begins as atoms above resonance, and the laser frequency is adjusted so that $\Delta$ changes adiabatically from positive to negative, then the system will follow the ground state as it evolves from atoms to stable molecules. Since the two-photon coupling $\chi$ sets the frequency scale, a change of laser frequency is adiabatic if $|\Delta_f-\Delta_i|\alt\chi$ for $(t_f-t_i)\alt\chi^{-1}$. Hence we use $\Delta(t)=\Delta_i\pm(\chi^2/10)t$. On a practical note, the change in the two-photon detuning $\Delta$ is effected by changing the frequency of laser~2, since the condition $\delta=100\Gamma_0$, which mitigates decay, fixes the frequency of laser~1.

\begin{figure}[t]
\centering
\includegraphics[width=8.5cm]{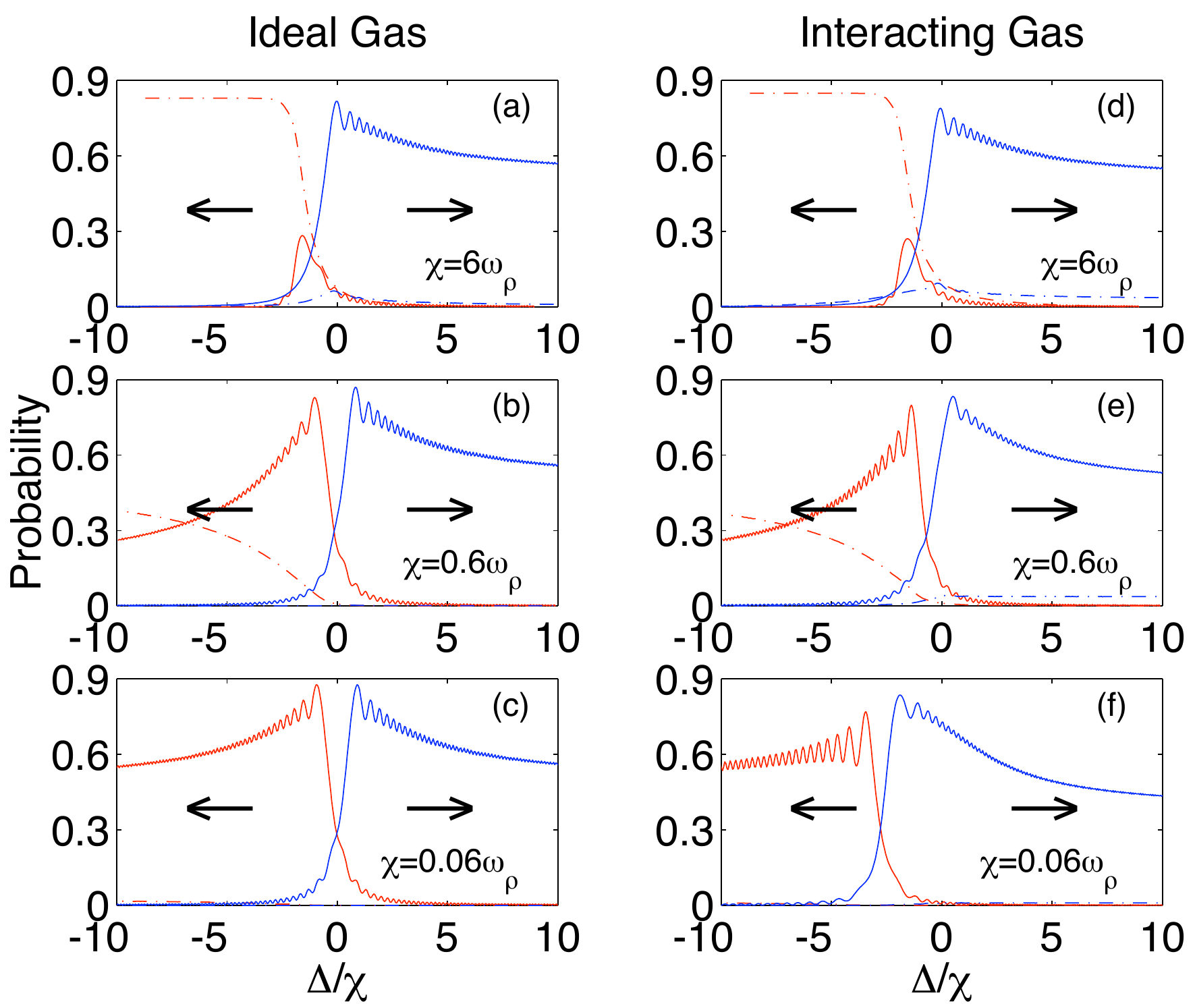}
\includegraphics[width=8.25cm]{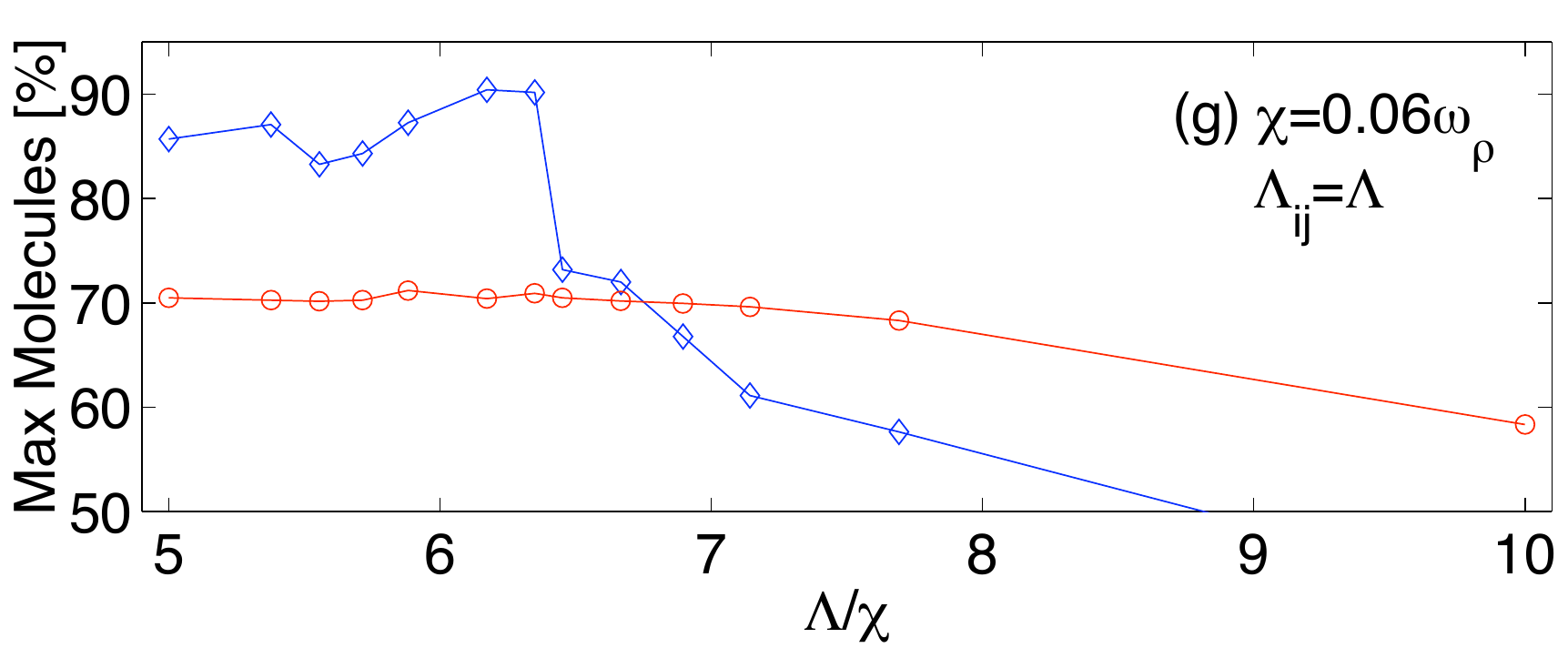}
\caption{(color online)~Probability vs. two-photon detuning for ideal~(a-c) and interacting~(d-f) gases, where the solid (dashed) lines are molecules (dissociated pairs), initial atoms are not shown, and the left arrows/red lines (right arrows/blue lines) denote $\Delta_{\rm initial}>0$ $(<0)$. (g) Maximum molecular percentage vs. collision strength for weak coupling, where the blue diamonds (red circles) denote $\Delta_{\rm initial}<0$ ($\Delta_{\rm initial}>0$).}
\label{SWEEPS}
\end{figure}

Basic adiabatic following, and the role of photodissociation, is illustrated for an ideal gas in \mbox{Figs.~\ref{SWEEPS}(a-c)}, where solid (dashed) lines are molecules (atom pairs) and initial atoms are not shown. For $\Delta_{\rm initial}<0$, as discussed above, the system follows the ground state as it evolves from atoms to molecules, independent of the atom-molecule coupling [Figs.~\ref{SWEEPS}(a-c), right arrows, blue curves]. For $\Delta_{\rm initial}>0$ at strong coupling [Fig.~\ref{SWEEPS}(a), left arrow, red curves], molecules form near resonance, but quickly turn into pairs since photodissociation is energetically favorable for $\Delta<0$. The weak coupling results [Fig.~\ref{SWEEPS}(c)] are independent of ${\rm sgn}(\Delta_{\rm initial})$, since photodissociation produces only a tiny fraction of pairs [barely visible in Fig.~\ref{SWEEPS}(c)] on the given timescale. 

Adding collisions makes little difference for strong and moderate coupling [Figs.~\ref{SWEEPS}(d) and~(e), respectively]. For weak photodissociation [Fig.~\ref{SWEEPS}(f)], collisions shift the two-photon resonance well off zero detuning, and lead to an asymmetry with respect to ${\rm sgn}(\Delta_{\rm initial})$. Atom-molecule conversion takes place largely near resonance within $\Delta\approx\pm\chi/2$, which for weak coupling translates into a rate $R\approx\chi/10\ll\omega_\rho$ and a timescale of about 0.5~ms, far below the rate limit~\cite{JAV02,MCK02}. We examine the role of collisions further by setting $\Lambda_{ij}=\Lambda$ and fixing $\chi=0.06\omega_\rho$. For increasing $\Lambda$, the conversion efficiency maximizes for $\Delta_{\rm initial}<0$, and conversion for $\Delta_{\rm initial}>0$ is relatively unaffected [Fig.~\ref{SWEEPS}(g)]. Compared to $\Lambda_{\rm max}/\chi\approx5$ in Fig.~\ref{SWEEPS}(f), conversion $\sim90\%$ survives an increase in the strength of elastic collisions $\sim30\%$. 

Finally, extrapolating to heavier systems should be possible. Recalling $\Lambda_{\rm max}/\chi\propto(a_{\rm max}/\mu_{\rm max})\sqrt{\mu_{12}/I_1}\,$, heavier particles will require less photoassociation intensity $I_1$ to satisfy weak coupling and maintain high efficiency. On the other hand, instead of reducing $I_1$, heavier particles allow high efficiency for a possibly larger s-wave scattering length $a_{\rm max}$.

{\em Four-Laser Scheme.}--We also consider the four-laser scheme illustrated in Fig.~\ref{FEWL}(c), where transitions to stable molecules occur through an intermediate, vibrationally-excited molecular state in the ground electronic manifold. The photoassociation laser still converts atom pairs in the state $|a_1,a_2\rangle$ into electronically-excited molecules in state $|e\rangle$, but the second laser now converts molecules in the state $|e\rangle$ into electronically-stable, vibrationally-excited molecules in state $|v\rangle$. A third laser then converts molecules in $|v\rangle$ into molecules in a second electronically-excited state $|e_2\rangle$, and a fourth laser converts molecules in state $|e_2\rangle$ into molecules in the stable state $|g\rangle$. The laser 1~and~2 couplings and detunings are as defined previously, whereas the laser 3~(4) coupling is $\Omega_{3(4)}$, the detuning of laser 3 from the $|v\rangle\leftrightarrow|e_2\rangle$ transition is $\delta_2$, and the detuning of lasers 3~and~4 from the two-photon transition $|v\rangle\leftrightarrow|g\rangle$ is $\Delta_{02}$. Spontaneous decay from the states $|e\rangle$ and $|e_2\rangle$ is taken to occur at the same rate $\Gamma_0$.

When laser 1 is off resonant from the transition $|a_1,a_2\rangle\leftrightarrow|e\rangle$, and laser 3 is off resonant with the $|v\rangle\leftrightarrow|e_2\rangle$ transition, then two-photon transitions $|a_1,a_2\rangle\leftrightarrow|v\rangle$ and $|v\rangle\leftrightarrow|g\rangle$ will dominate, as illustrated in Fig~\ref{FEWL}(d). The basic model~\eq{EQM} is then generalized to read 
\bml
\bea
i\dot{a}_1 &=& \omega_1 a_1 -\half\chi_1 a_2^* b, 
\\
i\dot{a}_2 &=& \omega_2 a_2 -\half\chi_2 a_1^* b,
\\
i\dot{b} &=&  (-\Delta_1+\omega_3-i\Gamma_1/2) b 
  -\half\chi_1 a_1a_2 -\chi_2 g
\nonumber\\&&
  -\frac{\chi_1}{8\pi^2\omega_\rho^{3/2}}\!\int\!d\varepsilon\sqrt{\varepsilon}\,f(\varepsilon)A(\varepsilon),
\\
i\dot{g} &=& [-(\Delta_1+\Delta_2)+\omega_4-i\Gamma_2/2]g-\chi_2 b,
\\
i\dot{A}(\varepsilon) &=& \varepsilon A(\varepsilon)
    -\chi\,f(\varepsilon)b,
\eea
\label{TWOSTEP_EQM}
\eml
where $b$ is the amplitude for the $|v\rangle$ molecules. The primary two-photon parameters are the same as Eqs~\eq{EQM}, but relabeled with subscript~1, and the secondary two-photon parameters are $\chi_2=\Omega_3\Omega_4/\delta_2$, $\Delta_2=\Delta_{02}+\Omega_4^2/\delta_2$, and $\Gamma_2=(\Omega_4/\delta_2)^2\Gamma_0$. Here the frequency of laser 1~(3) is fixed such that $\delta_{1(2)}=100\Gamma_0$, so that adiabatic following due to changing $\Delta_{1(2)}$ is effected by changing the frequency of laser 2~(4). The $\omega_i$ are again determined by $\Lambda_{ij}\propto\rho a_{ij}$, and collision-induced vibrational relaxation~\cite{BAL97} of molecules in $|v\rangle$ is included with imaginary scattering lengths estimated from $^{87}$Rb~\cite{WYN00,WIN05}.

\begin{figure}[t]
\centering
\includegraphics[width=8.5cm]{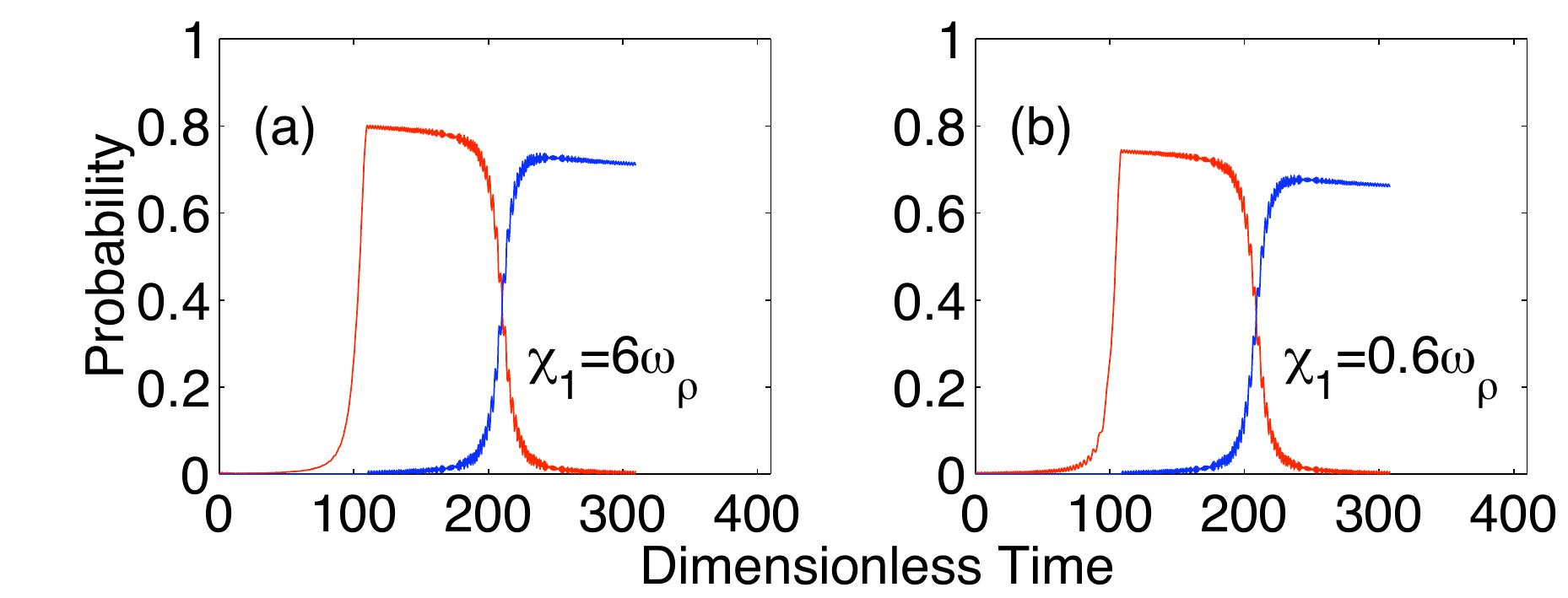}
\caption{(color online)~Molecular probability vs. dimensionless time for two-step adiabatic following, where the dimensionless time is $\tau_i=\chi_it$. In the first step, photoassociation creates vibrationally-excited molecules, which are converted into ground state molecules in the second step. The parameters for the photoassociation steps (red) are as in Fig.~\ref{SWEEPS}(d,e), respectively. The parameters for the second steps (blue) are fixed at $\Omega_3=10\Omega_4=\delta_2=100\Gamma_0$, so that $\chi_2/\Gamma_2=1000$.}
\label{TWO_STEP}
\end{figure}

Now a slow change in $\Delta_1$ converts atoms to vibrationally-excited molecules, and a subsequent slow change in $\Delta_2$ converts the vibrationally-excited molecules into stable molecules, where $\Delta_i=-|\Delta_{\rm initial}|+|\dot\Delta_i|t$ and $|\dot\Delta_i|=\chi_i^2/10$ is slow. The best yield is about 75\% [Fig.~\ref{TWO_STEP}(a)], but requires an impractically strong photoassociation laser, $I_1=14.2$~kW/cm$^2$, to compete against vibrational relaxation, even for low condensate density $\rho=10^{12}$~cm$^{-3}$. Moderate coupling [Fig.~\ref{TWO_STEP}(b)], requires less intensity, $I_1=1.42$~kW/cm$^2$, but the yield drops to about 65\%. Here atom-molecule conversion takes about 0.3~ms. Weak coupling, $\chi=0.06\omega_\rho$ (not shown), only requires $I_1=150$~W/cm$^2$, but only yields 35\% molecules.

{\em Conclusion.}--We have shown that, for practical laser intensities, a quantum degenerate gas of stable $^7$Li-Na molecules can be created with two-photon photoassociation using adiabatic following. Transitions directly to the stable molecular state require the least photoassociation intensity, about $30$~W/cm$^2$, and are thus more feasible than transitions via an intermediate, electronically-excited state. This requisite intensity decreases for heavier particles and, moreover, efficient low-intensity conversion is robust against reasonable increases in the strength of elastic s-wave collisions. Hence, although $^7$Li-Na was used as an example, the method should be feasible for heavier species, heteronuclear or homonuclear, and most likely for other statistics as well.

This work supported by the NSF (PHY-00900698).

\commentout{Crucially, $\chi/\Lambda_{\rm max}\propto\sqrt{(\mu_{\rm max}^2/\mu_{12})I_1}$, so that heavier particles require less photoassociation laser intensity to satisfy weak coupling.}

\end{document}